\documentstyle[12pt,epsf,epsfig,wrapfig]{article}
\begin{document}

\begin{center}
{\bfseries DIRECT MEASUREMENT OF $\Delta{G}/G$ AT COMPASS}

\vskip 5mm
Y. Bedfer (on behalf of the COMPASS collaboration)

\vskip 5mm
{\small
{\it
CEA/DAPNIA, Saclay and CERN, Geneva.
}\\
 {\it
E-mail: Yann.Bedfer@cern.ch
}}
\end{center}

\vskip 5mm
\begin{abstract}
 The gluon polarization $\Delta{G}/G$ is the key to a further clarification of the spin structure of the nucleon. The COMPASS collaboration at CERN has set out to undertake the direct determination of this quantity. It accesses the gluon distribution via the photon-gluon fusion process ($PGF$) in scattering polarized muons off a polarized deuteron target. And it explores three different channels to tag the $PGF$: open charm production and high transverse momentum (high $p_T$), in either electroproduction ($Q^2\,>\,1$~GeV$^2$) or quasi-real photoproduction ($Q^2\,<\,1$~GeV$^2$).

 The high $p_T$ quasi-real photoproduction channel yields the most precise measurement. The result indicates that the gluon polarization is small. I describe its experimental aspects and its theoretical framework, based on PYTHIA.  And I report on the preliminary results obtained in the other two channels, and on the prospects for future analysis and data taking.
\end{abstract}

\vskip 8mm 

\section{Introduction}
 The spin 1/2 of the nucleon can be decomposed as follows
 \[ \frac{1}{2} = \frac{1}{2}\Delta{\Sigma} + \Delta{G} + L_z ~~,\]
where the right hand side terms designate the contributions of the spin of the quarks, the spin of the gluons and the angular momentum of the quarks and gluons, respectively.

 In the recent years much effort has been put in determining $\Delta{\Sigma}$.
 This quantity can be derived from the measurement of the spin dependent 
structure function $g_1$ by polarized inclusive deep inelastic lepton-nucleon
 scattering ($DIS$) experiments. Measurements were carried out at CERN, SLAC, DESY and JLab. COMPASS itself is presently carrying on with this program. The
 results lead to the conclusion that $\Delta\Sigma$ is surprisingly small,
 significantly smaller than predicted by the Ellis-Jaffe sum rule~\cite{Ellis:1973kp} for
 example. An evaluation of $\Delta\Sigma$ taking into account the latest data points measured by COMPASS is presented by R. Windmolders at the SPIN05 workshop~\cite{RW}. 

 A solution to the problem was put forward in 1988~\cite{Efremov:1988zh,Altarelli:1988nr,Carlitz:1988ab}.
 It involves a leading order contribution to the polarized $DIS$ cross-section originating from the axial anomaly of QCD, $\alpha_s/2\pi~\Delta{G}$, which is anomalous in the sense that it does not vanish in the asymptotic limit:
in leading order evolution $\Delta{G}$ grows with $ln \, Q^2$ whereas
$\alpha_s$ is inversely proportional to  $ln \, Q^2$. This anomalous gluon
contribution introduces some freedom in the definition of $\Delta{\Sigma}$,
but it can in any case reconcile polarized $DIS$ data with QCD predictions given a large
enough, positive, value for the first moment of $\Delta{G}$~\cite{Cheng:1996jr}.

 The unpolarized gluon distribution, $G$, can be determined from the dependence
of the inclusive $DIS$ cross-section upon $Q^2$. In the polarized case,
however, $DIS$ data cover too small a range in $Q^2$ for this method to
significantly constrain $\Delta{G}$.

\begin{wrapfigure}[21]{R}{7cm}
\begin{center}
\mbox{\epsfig{figure=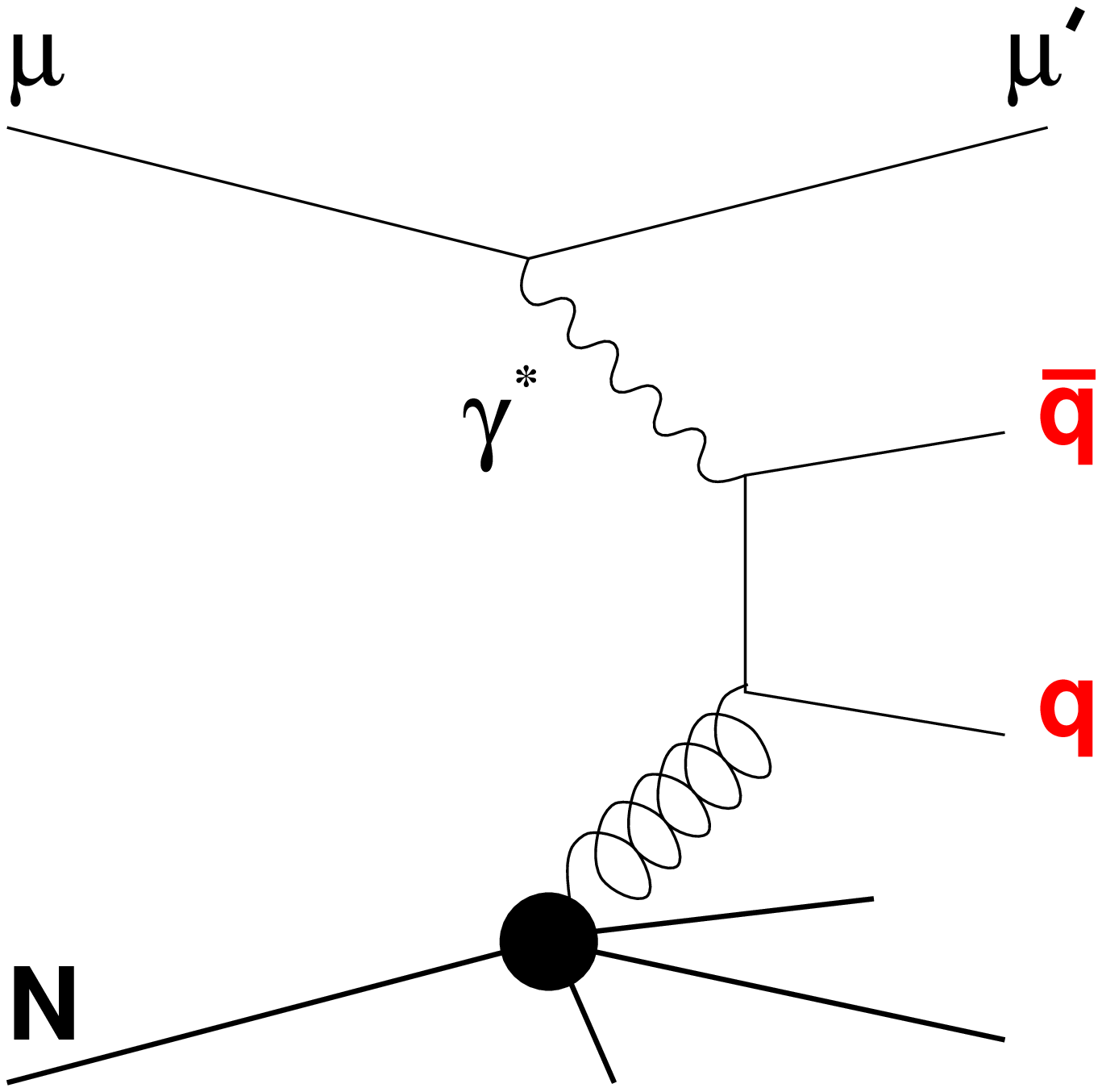,width=6cm}}
{\small{\bf Figure 1.} Photon-gluon fusion process in a muon nucleon scattering event.}
\end{center}
\end{wrapfigure}
 A direct measurement of the gluon polarization, $\Delta{G}/G$, is therefore the most promising way to further clarify the nucleon's spin puzzle. Several experimental projects have recently been started to carry it out, using different approaches. In COMPASS, we access the gluon distribution via the photon-gluon fusion ($PGF$) process (depicted on  Fig.~1), whereby a virtual photon couples to a gluon via a $q\bar{q}$ pair. And we consider two
different selections of the $PGF$: open charm production and high $p_T$ hadron
production.
The two share a set of common features. Both have been successfully used to directly measure the
unpolarized gluon distribution at the HERA collider experiments (with the
difference that high $p_T$ refers to the production of jets there).
And for both, the QCD scale can be set irrespective of
$Q^2$, by the charm mass and
the $p_T$ cut, respectively. But the two selections represent diametrically opposed trade-offs  between the conflicting requirements of statistics and purity. Open charm is the purest. It provides a model-independent access to $\Delta{G}/G$ and for this reason remains our golden channel. I will present it first.
And I will present next the high $p_T$ case, which we subdivide into two sub-cases depending upon the $Q^2$ of the exchanged photon. I start with some experimental essentials.

\section{Experimental essentials}
 The COMPASS spectrometer is described in details in~\cite{Mallot:2004gk}.

 Its experimental setup was designed to allow a precise determination of
asymmetries. An important point in this respect, is the control of fake asymmetries. We achieve it thanks to the simultaneous measurement of both orientations of
target spin in two oppositely polarized target cells, $u$ and $d$, and
to a frequent reversal of spin orientations, so that fluctuations in acceptance and incident
muon flux cancel out in the formula for the counting asymmetry $A$:
\begin{equation}
 A = \frac{1}{2}
  \left ( \frac{ N_u^{\Uparrow\uparrow} - N_d^{\Uparrow\downarrow} }{ N^{\Uparrow\uparrow}_{u} + N_d^{\Uparrow\downarrow} }+
  \frac{ N_d^{\Uparrow\uparrow} - N_u^{\Uparrow\downarrow} }{ N_d^{\Uparrow\uparrow} + N^{\Uparrow\downarrow}_{u} } \right)~~,
\label{A_RAW}
\end{equation}
where $\Uparrow\uparrow$ and $\Uparrow\downarrow$
denote the two spin configurations. (Note that weighted asymmetries are used instead of~(\ref{A_RAW}) in all calculations presented below.)

 The cross-section helicity asymmetry, $A_{\parallel}$, is related to the counting asymmetry by factors describing the polarization of the incoming particles, $P_b$ for the beam ($\sim$76\%), $P_t$ and $f$ for the target polarization ($\sim$50\%) and for the,
 process dependent, dilution factor ($\sim$40\%). It is best expressed as $A_{\parallel}/D$,
\begin{equation}
 A_{\parallel}/D = A~/~( P_b \times P_tf \times D )~~,
\label{A_PARALLEL}
\end{equation}
where one takes also into account a kinematical factor, $D$, describing the polarization transfer from the muon to the photon. $D$ is process dependent and typically averages to $\sim$60\%.

 During its three years of running from 2002 to 2004, the experiment has accumulated $\sim$2.4~$fb^{-1}$ of data with its target polarized longitudinally. The results presented here correspond to about half of these data.

\section{Open Charm}
  This channel was discussed by many authors~\cite{Watson:1981ce, Gluck:1988uj} as a good candidate to access $\Delta{G}/G$. Since there is no or only a small intrinsic charm in the nucleon, diagrams with an incoming charm quark do not contribute and $PGF$ enters alone in leading order.

 In COMPASS we tag open charm, and hence $PGF$, by the reconstruction of a $D^o$ meson. \begin{figure}[h]
\begin{center}
\mbox{\epsfig{figure=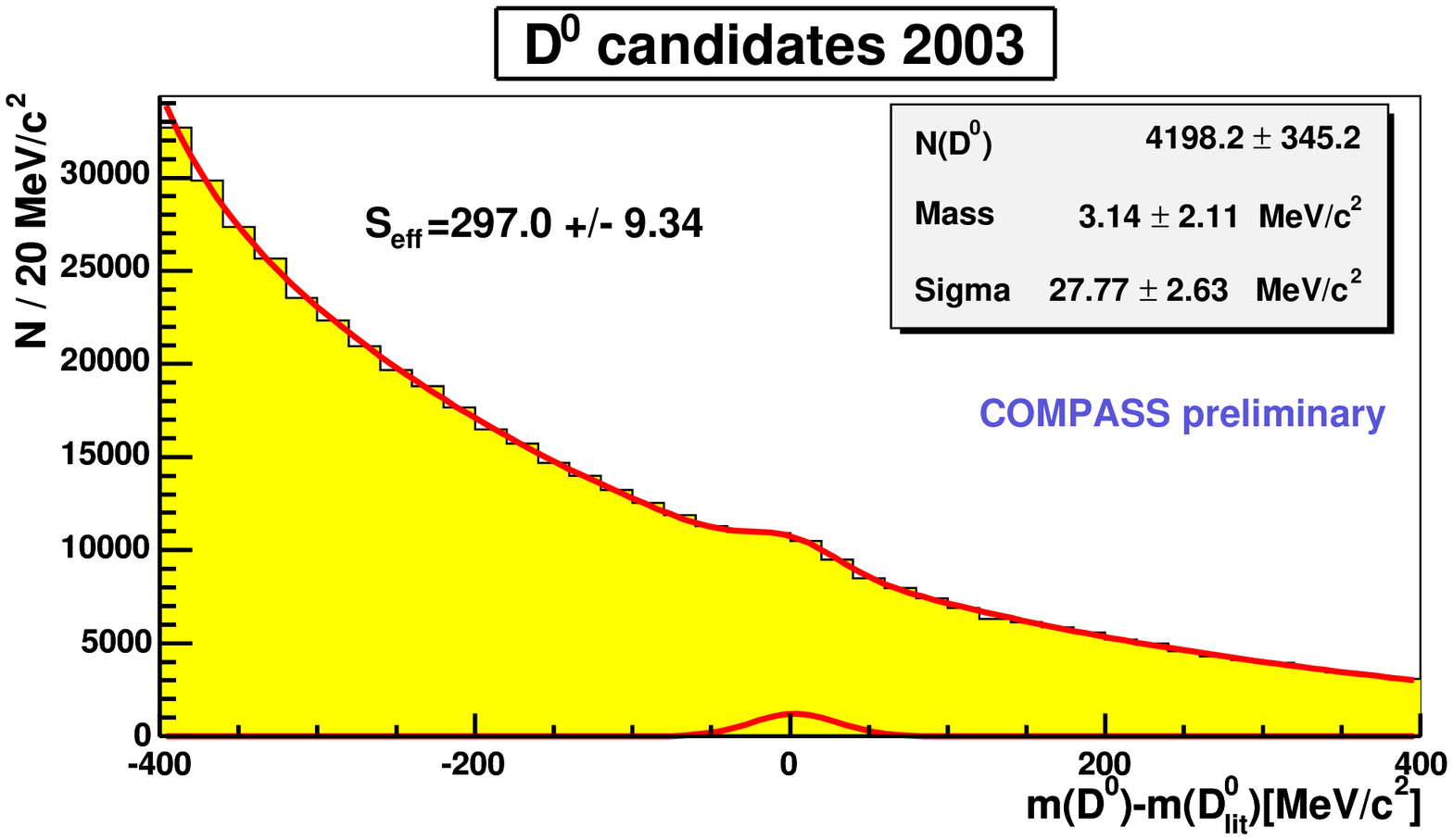,width=.65\linewidth}}
\end{center}
\begin{center}
\mbox{\epsfig{figure=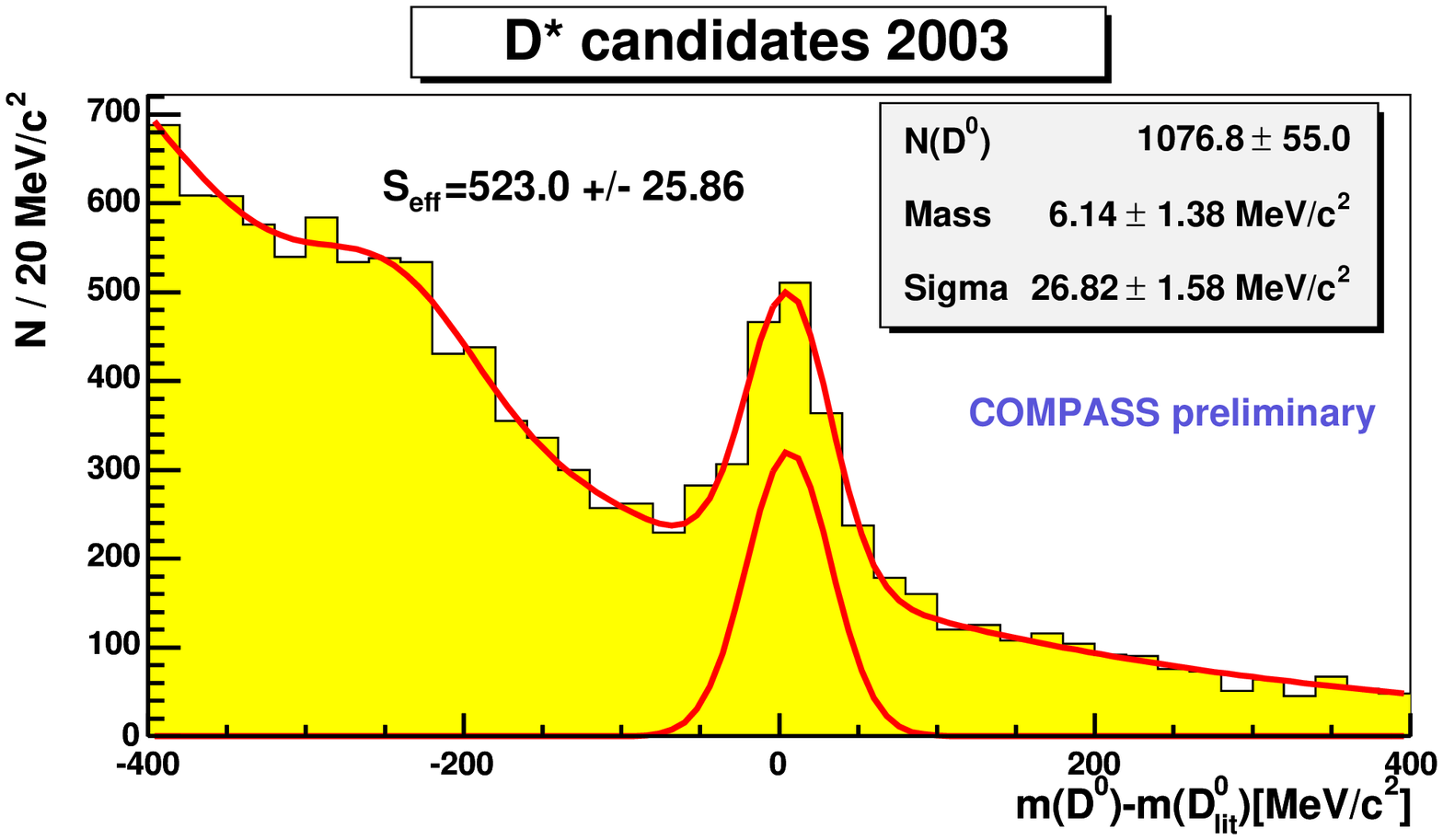,width=.65\linewidth}}
\end{center}
{\small{\bf Figure 2} $D^o$ peak in the $K\pi$ invariant mass distribution for all events (top) and $D^*$-tagged events (bottom) for 2003 ({\it i.e.} $\sim$1/3 of total) data. $S_{eff}$ is the effective number of events
 $S_{eff} = S^2/(S+B)$ where $S$ and $B$ are signal and background counts. A bump shows up at low mass in the $D^*$ case, attributable to $D^o \rightarrow K\pi\pi^o$. It is not included in the $S$ count.}
\end{figure}
The $D^o \rightarrow K\pi$ decay channel is used, with $K$ and $\pi$ identified in the RICH. The main
difficulty lies in the associated combinatorial background.
This is a major concern in our experiment, where the vertex resolution is
not sufficient to resolve  the decay vertex from the primary vertex, because of the thickness
of the target.

 Special care is therefore taken to optimize the use of the data.
First, the favorable cases when the $D^o$ comes from a
$D^*\rightarrow D^o\pi$ decay are counted separately, {\it cf.} Fig.~2.
Secondly, a weighting procedure is applied for the
 derivation of $\Delta G/G$:
\begin{equation}
  \Delta G/G = \frac{1}{P_T}~~\frac{\sum_i^{\Uparrow\uparrow} w_i - \sum_i^{\Uparrow\downarrow} w_i}{\sum_i^{\Uparrow\uparrow} w_i^2 + \sum_i^{\Uparrow\downarrow} w_i^2}~~~~~~~~~~w_i = \frac{\langle f\,P_b\,a_{LL} \rangle_i}{(1+B/S)_i}
\label{DGG_OC}
\end{equation}
where $a_{LL}$ is $PGF$'s analyzing power and $S$ and $B$ are the signal and
background counts.

 A Monte Carlo simulation of the experiment is used to calculate estimates of all quantities that cannot be directly computed from the hadron level kinematics, such as $a_{LL}$ and $x_g$, the momentum fraction carried by the gluon. It is based on the event generator AROMA~\cite{Ingelman:1996mv}, which uses the $PGF$ matrix element to generate charmed hadrons. $a_{LL}$ is first parameterized as a function of the transverse momentum and energy fraction of the reconstructed charmed meson, and of a kinematical factor, $D$, describing the polarization transfer to the virtual photon, so that it can be estimated on an event by event basis in equation~(\ref{DGG_OC}).

 In these conditions, a preliminary analysis, bearing on $\sim$1/2 of the data, yields:
\begin{center}
 $\Delta G/G = -1.08 \pm 0.73(stat.)$ ~~~~~at $x_g = 0.15 \pm 0.08\, RMS$.
\end{center}

 This means a clear lack of precision, even when one takes into account all of the available statistics. Obviously more data are needed. COMPASS is going to resume data taking in 2006, with an improved experimental apparatus. Several upgrades, targeting three important aspects of the $D^o$ selection will be introduced:
\begin{itemize}
\item
An enlarged angular acceptance, thanks to a new target superconducting solenoid  (with an aperture of 180~mrad instead of 70~mrad), and a better coverage of large angles by drift chambers.
\item
A better particle ID, thanks to a faster RICH detector, able to clear away the high level of background generated by the halo of our muon beam.
\item
Electromagnetic calorimetry, that will allow us to explore $D^o$ decays with a $\pi^o$ in the final state, mainly $D^o \rightarrow K\pi\pi^o$ with a $BR$ of 13\% against 3.8\% for $D^o \rightarrow K\pi$.
\end{itemize}

\section{High $p_T$}
The alternative approach to select $PGF$ is to require hadron production with a high transverse momentum  with respect to the virtual photon~\cite{Bravar:1997kb}. This suppresses leading order $\gamma^*q \rightarrow q$ events, where the fragmenting quark goes into the direction of the photon. However the suppression is not perfect. And other competing second order processes have to be taken into account. These include the QCD Compton process, and, mainly at low $Q^2$, processes involving resolved photons. In these conditions, and limiting oneself to leading order, where the different processes are well separated, the cross-section helicity asymmetry reads:
\begin{equation}
  A_{\parallel} = R_{PGF} \: a^{PGF}_{LL} \: \Delta G/G \; + \; A_{background}~~,
\label{HPT_1}
\end{equation}
where $R_{PGF}$ is the fraction of $PGF$ events, $a^{PGF}_{LL}$ is $PGF$ analyzing power and $A_{background}$ describes the contribution of the background processes. In order to retrieve $\Delta G/G$, $R_{PGF}$, $A_{background}$ and the parton level kinematics defining $a^{PGF}_{LL}$ need be determined by a simulation of the experiment. In COMPASS, we resort to the event generators LEPTO~\cite{Ingelman:1996mq} and PYTHIA~\cite{Sjostrand:2003wg} to do the computation, which introduces a model dependence in our measurement. LEPTO covers the electroproduction regime at $Q^2 \, > \, 1$~GeV$^2$ and PYTHIA the quasi-real photoproduction regime at $Q^2 \, < \,1$~GeV$^2$. In both cases, the leading order approximation is consistently applied, in the choice of PDFs (we use GRV98~\cite{Gluck:1998xa} and GRSV2000~\cite{Gluck:2000dy} at leading order), in the calculation of the analyzing powers and by turning off parton showers in JETSET. And for both, the event selection follows a same path. Two hadrons are required, with $p_T> \, 0.7$~GeV and $\Sigma \, p_T^2 \, > \, 2.5$~GeV$^2$. Contributions from target fragmentation and resonances are excluded by requiring $x_F \, > \, 0.1$ and the invariant mass of the two-hadron system to be larger than $1.5$~GeV. But the two cases are attractive in their own right: $Q^2<1$~GeV$^2$ yields much higher statistics, a factor 10, but it also involves more model dependence.

\subsection{High $p_T$, $Q^2 \, < \,1$~GeV$^2$}
 The details of the measurement are available in~\cite{Ageev:2005pq}. I recall here the main aspects.

 PYTHIA provides a complete coverage of photon-nucleon interactions at low $Q^2$~\cite{Friberg:2000ra}. We checked that it reproduces reasonably well our own data, by propagating generated events through a GEANT model of the spectrometer and processing them through the full reconstruction and analysis chain. A good agreement was readily obtained for the kinematical variables $Q^2$ and $y$. A single parameter, the width of the intrinsic $k_T$ of the partons in the photon, had to be adjusted (from 1 to 0.5~GeV) to fit the longitudinal and transverse momentum distributions of the two hadrons. As an example, Fig.~3 shows the agreement obtained for  the $p_T$ distribution of one of the hadrons.  The relative contributions, $R$, of the various processes thus generated are shown on Fig.~4.

\begin{figure}[h]
\begin{center}
\mbox{\epsfig{figure=pt0.epsi,width=.65\linewidth, angle=-90}}
\end{center}
{\small{\bf Figure~3} Comparison between data and Monte Carlo. The $p_T$ distributions for the hadron with highest $p_T$ are displayed, superimposed, for the two of our triggers. The bottom plots show the ratios: data divided by Monte Carlo.}
\end{figure}

\begin{figure}[h]
\begin{center}
\mbox{\epsfig{figure=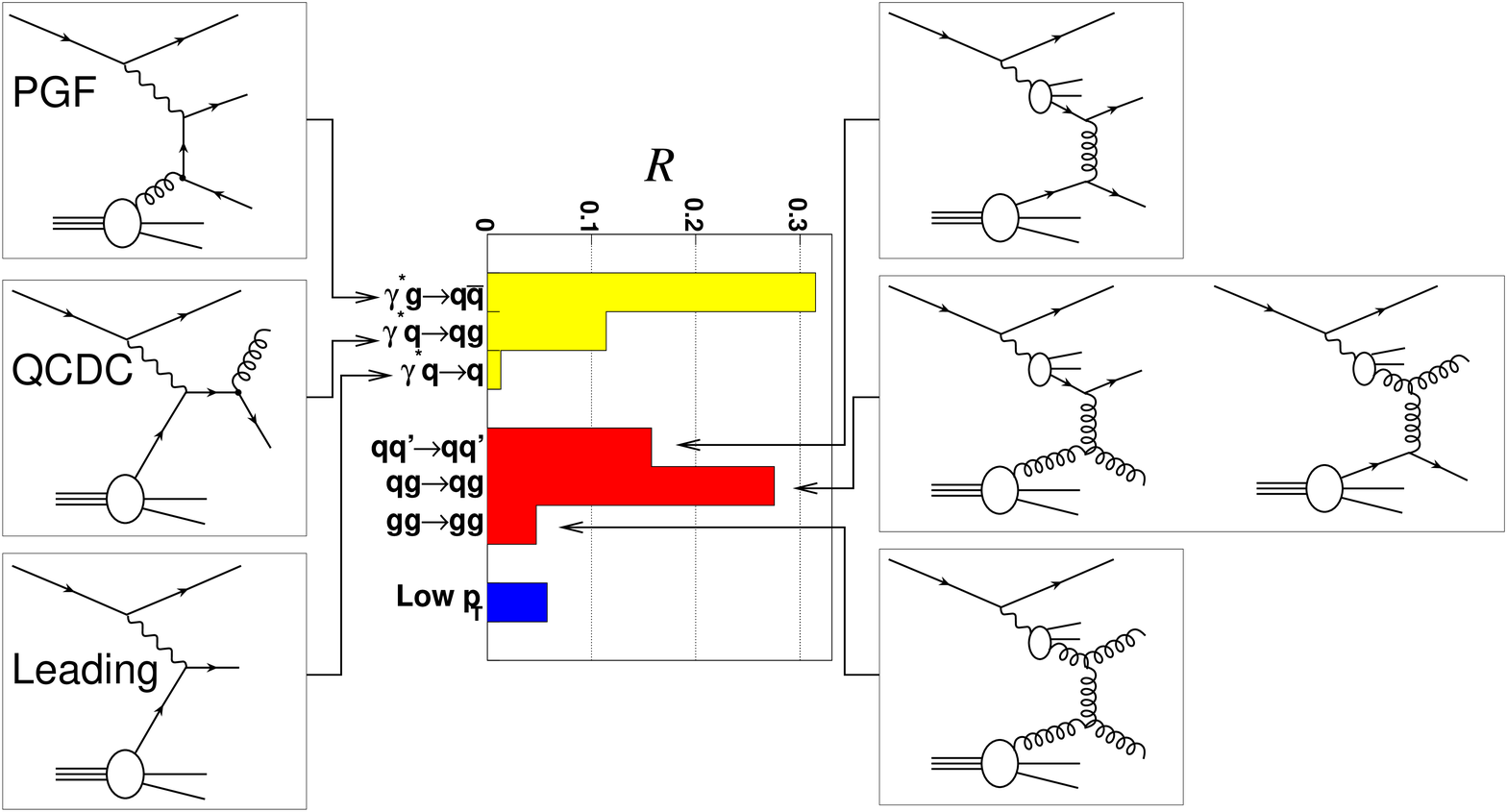,width=.95\linewidth}}
\end{center}
{\small{\bf Figure 4} Relative contributions $R$ of the dominant PYTHIA processes to the Monte Carlo sample of high $p_T$ events at $Q^2\,<\,1$~GeV$^2$. Left: direct processes, right: resolved photon processes. Longitudinal photons, as well as some minor resolved photon contributors, are not shown.}
\end{figure}

 To compute the asymmetry of a given process, it is mandatory to find a hard scale allowing the factorization of the cross-section into a hard partonic cross-section calculable perturbatively and a soft parton distribution function. In our case, this hard scale is to be looked for in the $p_T$ of the partons. Thanks to the hadron $p_T$ cut, for most events, one parton at least turns out to fulfill the condition. However this is not true for leading order $\gamma^*q \rightarrow q$ events, and for some other events which PYTHIA classifies as ``Low-$p_T$''. In these conditions, we can expand equation~(\ref{HPT_1}) as follows:
\pagebreak
 \[ \langle \frac{A_{\parallel}}{D} \rangle \: = R_{PGF} \:
  \langle \frac{a_{LL}^{PGF}}{D} \rangle \: \frac{\Delta G}{G} \; + \;
  R_{\small{QCDC}} \: \langle \frac{a_{LL}^{QCDC}}{D} \, A_1 \rangle \; + \]
\begin{equation}
 \sum_{f,f^{\gamma} = u,d,s,\bar{u},\bar{d},\bar{s},G} \! R_{ff^{\gamma}} \;
  \langle a_{LL}^{ff^{\gamma}} \, \frac{\Delta f}{f} \, \frac{\Delta f^{\gamma}}{f^{\gamma}} \rangle + \: R_{Leading} \: A_{Leading} \; + \; R_{Low-p_T} \: A_{Low-p_T}~~,
\label{HPT_2}
\end{equation}
where the QCDC term makes use of the inclusive photon-deuteron asymmetry, $A_1$, and the resolved photon term makes use of the PDFs in the deuteron, $f$ and $\Delta{f}$, and in the photon, $f^{\gamma}$ and $\Delta{f^{\gamma}}$, and where $D$ is a kinematical factor which approximates the amount of polarization transfered from the muon to the photon. The last two terms, corresponding to the leading and low $p_T$ processes, cannot be expanded, since there is no hard scale allowing to factorize their cross-section factor. However, the asymmetry for this kind of events is known to be small, as indicated by measurements at low $Q^2$~\cite{Adeva:1999pa}, and they account for only 7\% of the high $p_T$ sample, {\it cf.}~Fig.~4. Therefore, they are neglected. All the PDFs appearing in the remaining terms, apart from the targeted $\Delta{G}$, have been measured, except the polarized distributions of the photon $\Delta{f^{\gamma}}$. However, each of these is bound by a minimal and a maximal saturation scenarios, as described in~\cite{Gluck:2001rn}\footnote{Our definition of the minimal scenario corresponds to a maximally negative polarization of the parton in the VMD part of the photon.}. We can therefore solve equation~(\ref{HPT_2}) to determine $\Delta{G}/G$ from the measurement of $\langle A_{\parallel}/D \rangle$. Note that the $PGF$ term is not the only one containing  $\Delta{G}/G$, the resolved photon term is also doing so. The two contributions have opposite signs and cancel each other to some extent: the $PGF$ term is equal to $-0.29\,\times\,\Delta{G}/G$ and the linear coefficient of $\Delta{G}/G$ in the resolved photon term is positive, ranging from +.012 to +0.078 depending upon which saturation scenarios are used.

 The Monte Carlo analysis outlined above introduces systematic uncertainties.  We have considered the following two sources: NLO effects and the tuning of PYTHIA. We have estimated their order of magnitude by repeating the analysis several times with modified Monte Carlo parameters. For the NLO effects, the renormalization and factorization scales were multiplied and divided by two and the parton shower mechanism was activated. For the other source, we have explored the parameter space, while maintaining a reasonable agreement between simulated and real data. The most critical parameter appeared to be the width of the intrinsic $k_T$ in the photon. 

 $A_{\parallel}/D$ can be obtained from equations~(\ref{A_RAW}),(\ref{A_PARALLEL}). In order to minimize the statistical error, we use instead the weighted method described in~\cite{Adams:1997tq}. With the high $p_T$ selection defined above, and the added condition $0.35\,<\,y\,<0.9$, we obtain:
\begin{center}
   $ \langle A_{\parallel}/D \rangle = 0.002 \pm 0.019(stat.) \pm 0.003(syst.)~~, $
\end{center}
where the systematic uncertainty accounts for false asymmetries, which were estimated using a sample of low $p_T$ events with much larger statistics.

 For the derivation of $\Delta{G}/G$, we first consider the minimum and maximum values obtained from equation~(\ref{HPT_2}) with the various saturation scenarios for the polarized PDFs of the photon mentioned above:
\pagebreak
\begin{center}
  $(\,\Delta{G}/G\,)_{min} =  0.016 \pm 0.068 (stat.) \pm 0.011 (exp.syst.) \pm 0.018 (MC syst.)$\\
  $(\,\Delta{G}/G\,)_{max} =  0.031 \pm 0.089 (stat.) \pm 0.014 (exp.syst.) \pm 0.052 (MC syst.)$
\end{center}
This leads to the central value:
\begin{center}
$ \Delta{G}/G = 0.024 \pm 0.089 (stat.) \pm 0.057 (syst.)$ ~~at~~ $x_g = 0.095^{+0.08}_{-0.04}$ ~and~ $ \mu^2 \simeq 3 \,$~GeV$^2$.
\end{center}
where the difference between the min and max values is included in the systematics and all systematics are added quadratically. $x_g$ and the average scale $\mu^2$ are both obtained from the Monte Carlo simulation.

\subsection{High $p_T$, $Q^2 \, > \, 1$~GeV$^2$}
 This time the analysis is only preliminary. It follows almost the same path as the one just described for the low $Q^2$ case. And I will only mention what singles it out.

 The $Q^2$ cut introduces several simplifications to equation~(\ref{HPT_2}). It provides a hard scale for all processes. This eliminates all purely soft processes previously classified as ``Low-$p_T$''. And it allows us to factorize the leading process term into a hard cross-section and a soft part, which we approximate by $A_1$ as for the QCDC term. Besides it reduces drastically the cross-section of the resolved photon processes, which are altogether neglected.

 Further simplification is obtained by restricting the analysis to the low Bjorken $x$ region, $x_B\,<\,0.05$, where $A_1$ for the deuteron is small~\cite{Ageev:2005gh}. Hence, the contributions of the QCDC and leading processes to the asymmetry become negligible.

  After all selection cuts, we obtain:
\begin{center}
$ \langle A_{\parallel}/D \rangle = -0.015 \pm 0.080(stat.) \pm 0.013(syst.) $
\end{center}
  The Monte Carlo simulation is performed using LEPTO, with a set of fragmentations parameters modified to fit the data. It yields $R_{PGF}\,\simeq\,34\%$, ~ $\langle\,a_{LL}/D\,\rangle\,\simeq\,-75\%$  and:
\begin{center}
$ \Delta G/G = 0.06 \pm 0.31(stat.) \pm 0.06(syst.)$ ~~at~~ $x_g = 0.15 \pm 0.08 \, RMS~~,$
\end{center}
where the systematical error accounts only for the experimental uncertainty.

\section{Conclusion and outlook}

 Our high $p_T$ results for the gluon polarization are plotted on Fig.~5. They are compared with previous direct measurements by SMC~\cite{Adeva:2004dh} and HERMES~\cite{Airapetian:1999ib}. The SMC measurement uses high $p_T$ events at $Q^2\,>\,1$~GeV$^2$, while the HERMES result is derived from data mostly at low $Q^2$. Both neglect resolved photon processes.

\begin{figure}[h]
\begin{center}
\mbox{\epsfig{figure=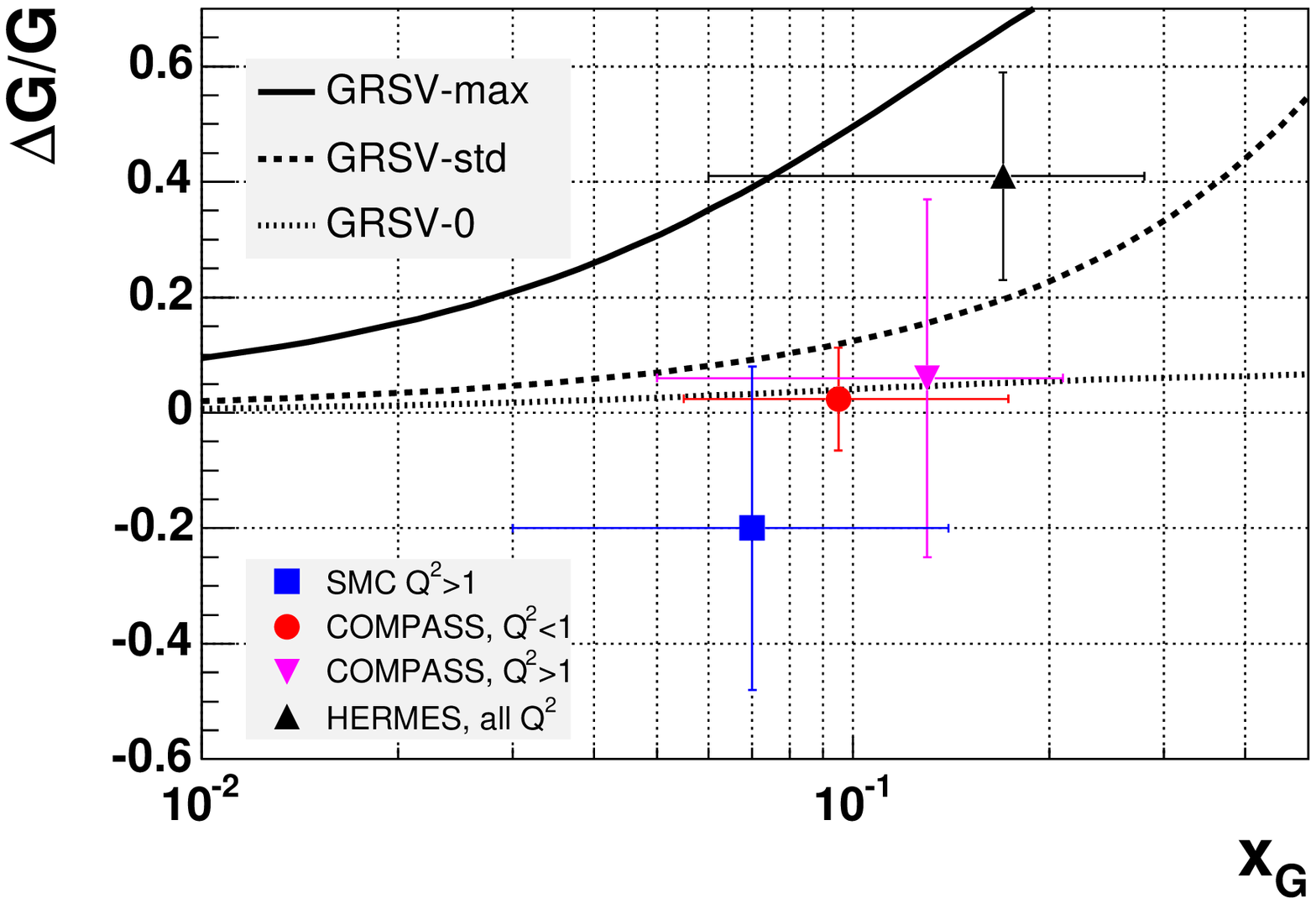,width=.95\linewidth}}
\end{center}
{\small{\bf Figure~\ref{DGG_GRSV}} Comparison of the $\Delta{G}/G$ measurements from COMPASS (the two high $p_T$ results are shown, that at $Q^2\,>\,1$~GeV$^2$ being preliminary), SMC~\cite{Adeva:2004dh} and HERMES~\cite{Airapetian:1999ib}. For each point, the vertical error bar includes the statistical uncertainty only and the horizontal bar represents the $x_g$ range. The curves show three parameterizations of $\Delta{G}/G(x_g)$ at $\mu^2\,=\,3$~GeV$^2$ from the NLO fits of~\cite{Gluck:2000dy}. (see text for details).}
\label{DGG_GRSV}
\end{figure}

 It is interesting to compare our most precise result, from high $p_T$ at low $Q^2$, with distributions of $\Delta{G}/G(x_g)$ obtained from QCD fits. Three such distributions, from GRSV2000~\cite{Gluck:2000dy}, are plotted on Fig.~5. They correspond to three hypotheses on the gluon polarization at an input scale of $\mu^2\,=\,0.40$~GeV$^2$: maximal polarization (max), best fit (std) and zero polarization (0). The distributions are then evolved to $\mu^2\,=\,3$~GeV$^2$, where their first moments are 2.5, 0.6 and 0.2, respectively. Our low $Q^2$ result clearly favors parameterizations with a low gluon polarization. Two other recent NLO fits~\cite{Hirai:2003pm, Leader:2005kw} (not shown), with first moment values at  $\mu^2\,=\,3$~GeV$^2$ ranging from 0.26 to 0.8, were also checked against our result at $x_g\,=\,0.095$ and found to be compatible within 1.5~$\sigma$. The anomaly mechanism mentioned in the introduction would require a first moment of $\Delta{G}$ of about 3 to explain the polarized $DIS$ data. The small value of $\Delta{G}/G$ from our measurement cannot rule out this possibility. However it makes it very unlikely.

  The two high $p_T$ results discussed above assume the validity of PYTHIA and LEPTO, respectively. The open charm channel would allow instead to make a model independent measurement. It still suffers from a lack of statistics but remains our golden channel. COMPASS resumes data taking in 2006, with an improved experimental apparatus, and will strive to complete the measurement.

 The high $p_T$ channels will also benefit from the new data. They will have to include first the already existing data which remain to be analyzed. Which will bring down the statistical precision to $\delta\Delta{G}/G = 0.22$ for the high $Q^2$ case and  $\delta\Delta{G}/G = 0.06$ for the low $Q^2$ case. For the latter, the aim will be to use the extra statistics brought by 2006 data to create two bins in $x_g$.

\end{document}